\documentclass[preprint,review,12pt]{elsarticle}

\usepackage{amssymb}
\usepackage{amsthm}
\usepackage{amsmath}
\usepackage{multirow}
\usepackage{booktabs}
\usepackage{graphicx}
\usepackage{epstopdf}
\usepackage{pdflscape}
\usepackage{amsthm}
\usepackage{amsmath}
\usepackage[english]{babel}
\usepackage{subfigure}
\usepackage{threeparttable}

\usepackage{algorithm,algorithmicx,algpseudocode}

\usepackage{tabularx}
\usepackage{graphicx}
\usepackage{caption}
\usepackage{threeparttable}
\usepackage[figuresright]{rotating}
\usepackage[toc,page,title,titletoc,header]{appendix} 

\makeatletter

\newcommand{\Rmnum}[1]{\expandafter\@slowromancap\romannumeral #1@}
\makeatother

\journal{XXX}

\begin{document}

\begin{frontmatter}
\title{A Lagrangian Inertial Centroidal Voronoi Particle method for dynamic load balancing in particle-based simulations}

\author{Zhe Ji}
\ead{zhe.ji@tum.de}

\author{Lin Fu}
\ead{lin.fu@tum.de}

\author{Xiangyu Y. Hu}
\ead{xiangyu.hu@tum.de}

\author{Nikolaus A. Adams}
\ead{nikolaus.adams@tum.de}

\address{Technical University of Munich, Department of Mechanical Engineering, Chair of Aerodynamics and Fluid Mechanics, 85748 Garching, Germany}

\begin{abstract}

In this paper we develop a Lagrangian Inertial Centroidal Voronoi Particle (LICVP) method to extend the original CVP method \cite{fu2017physics} to dynamic load balancing in particle-based simulations. Two new concepts are proposed to address the additional problems encountered in repartitioning the system. First, a background velocity is introduced to transport Voronoi particle according to the local fluid field, which facilitates data reuse and lower data redistribution cost during rebalancing. Second, in order to handle problems with skew-aligned computational load and large void space, we develop an inertial-based partitioning strategy, where the inertial matrix is utilized to characterize the load distribution, and to confine the motion of Voronoi particles dynamically adapting to the physical simulation. Intensive numerical tests in fluid dynamics simulations reveal that the underlying LICVP method improves the incremental property remarkably without sacrifices on other objectives, i.e. the inter-processor communication is optimized simultaneously, and the repartitioning procedure is highly efficient.

\end{abstract}


\end{frontmatter}


1\section{Introduction}
\label{S:1}

Large scale parallel computing is essential for a wide range of scientific applications. The objective of the domain decomposition method is to facilitate the algorithms to harness computational resources more efficiently \cite{devine2005new}. In Computational Fluid Dynamics (CFD), the configuration of discretization-elements (mesh or particle) may evolve in time \cite{schloegel2000graph}, which requires partitioning algorithms to reassign computational load to processors periodically, i.e., achieve dynamic load balance. Comparing to static partitioning, rebalancing needs to meet the same targets, e.g. load balance, locality, optimization for inter-processor communication, but also is subject to further constrains. The key aspect for rebalancing schemes is to minimumize partitioning modifications subject to small topology changes, i.e. incremental property, and to optimize the inter-processor communication at smallest cost \cite{hendrickson2000dynamic} \cite{fu2016novel}. The reader may refer to \cite{bulucc2016recent} and \cite{schloegel2000graph} for a comprehensive review regarding recent development of repartitioning approaches.

For particle simulations, the discretization-elements, e.g. SPH particles, move according to the corresponding dynamic system, which may cause deformation and relocation of computational sub-domains. To achieve the incremental property it is crucial to maximize the data reuse during rebalancing process, namely, the rebalancer should be aware of the details of the underlying system, and repartition the system incorporating the existing partition to achieve lower data redistribution cost \cite{walshaw1997parallel}. Another problem encountered in particle simulation is that in some cases, e.g. the dambreak problem \cite{adami2012generalized}, high velocity impact of a projectile into a plate \cite{zhang2017generalized}, the rotating disk problem in astrophysics \cite{hopkins2015new}, the computational load distributes anisotropically in the computational domain, and the configuration of discretization-elements varies rapidly. The skew alignment and rapid change of computational load may result in void space which requires zero computing resources. Such a scenario is more problematic for partitioning algorithms with respect to satisfying all constrains simultaneously \cite{hendrickson2000dynamic}\cite{meyerhenke2009graph}.

In our previous paper \cite{fu2017physics}, a CVP domain decomposition method based on physical analogy has been developed. The load balance target is achieved by solving a Voronoi Particle (VP) evolution model equation. Centroidal Voronoi Tessellation (CVT) is utilized for communication reduction by optimizing the compactness of partitioning sub-domains \cite{meyerhenke2009graph}. The CVP method is verified by various static partitioning tests, independently of the mesh-element type. Later, we have integrated the CVP method as dynamic partitioner and develop a new multi-resolution parallel framework for the SPH method \cite{zhe_ji_new}. An adaptive rebalancing criterion and monitoring system has been proposed to assess the imbalance during simulation and reassign equivalent load among all the processors. 

Although the CVP method inherently features load balance and communication reduction, there is no explicit treatment in our previous work regarding to the handling of the aforementioned additional difficulties that will encountered in dynamic partitioning. The objective of the current paper is to extend the CVP method and improve the performance of particle-based methods in the dynamic partitioning problems. Two concepts are proposed in this paper: a Lagrangian background velocity and an inertial-based partitioning strategy. The newly developed method is called Lagrangian Inertial CVP (LICVP) method. In the following two paragraphs the basic idea of LICVP method is introduced.
 
With the CVP method each Voronoi particle possesses physical properties either integrated from meshes enclosed within its Voronoi cell region or calculated from neighboring cells. Generally speaking, the CVP method can be viewed as a coarse-grained modeling of underlining mesh elements. Based on this observation we can introduce a background velocity to adevect partitioning generators, i.e. Voronoi paticles, between two consecutive rebalancing steps. If the background velocity is given properly, the positions of the Voronoi particles will be updated describing the evolution of the dynamic system and geometry variation, e.g. mass center, of local sub-domains. When the rebalancing subroutine is triggered, the updated positions of Voronoi particle are utilized as input and initial condition for the new partitioning result. Since the equilibrium is calculated globally, the target of load balance and communication reduction is guaranteed by the CVP method. Moreover, since the new partitioning diagram is calculated aware of the original partitioning, data reuse is improved, i.e. incremental property can be achieved.

To handle problems with skew-aligned computational load and large void space, we propose another extension of the CVP method. The idea is to constrain the motion of Voronoi particles according to the load distribution in space during the rebalancing procedure. Similarly with the Recursive Inertial Bisection (RIB) method \cite{simon1991partitioning}, we choose the inertia matrix to characterize the load distribution, such that the skewness of the computational load can be obtained accordingly. A splitting operator is proposed to take the trajectory of Voronoi particles as the input and outputs the resultant vector subject to a certain type of constraint. Moreover, an adaptive filter is proposed, which selects a proper constraint dynamically adapting to the development of the underlying particle system. The filter ensures restoration of the original CVP method when the computational load is distributed homogeneously. With this method Voronoi particles are insensitive of the load variation along the confined direction, thus the convergence and incremental property are improved. We refer the new method as Inertial CVP.

In this paper, we focus on a specific version of particle-based method, the Smoothed Particle Hydrodynamics method. The validation of our new algorithm is performed with the code we have developed previously \cite{zhe_ji_new}. The remaining of this paper is arranged as follows. In section \ref{S:2} the CVP method and imbalance monitoring strategy for dynamic load balancing is briefly reviewed. The concept and model equations of the Lagrangian Inertial CVP method are introduced in section \ref{S:3}. Numerical algorithms and boundary conditions are elaborated in section \ref{S:4}. Section \ref{S:5} gives five numerical tests to verify our method.

\section{Brief review of the CVP method}
\label{S:2}

We briefly review the model equations of the CVP method and the imbalance monitoring strategy from previous work \citep{fu2017physics}\cite{zhe_ji_new}.

\subsection{Model equations}
\label{S:2_1}

The key concept of the CVP method is to combine CVT \cite{okabe2009spatial} and Voronoi Particle dynamics (VP) to achieve high-level compactness of partitioning sub-domains and error-controlled load balance simultaneously. The equilibrium is calculated iteratively utilizing a two-step time integration scheme. The CVT diagram is constructed employing the Lloyd method \cite{lloyd1982least}\cite{du2006convergence}. The model equation for VP is 

\begin{equation}
\label{eq:VP_dynamics_EOM}
\textbf{a}_{i}^{vp}=\dfrac{d\textbf{v}^{vp}_{i}}{dt}=-\dfrac{\int_{\Omega_{i}}\bigtriangledown pd\sigma}{\int_{\Omega_{i}}\rho d\sigma}=-\dfrac{\int_{\partial\Omega_{i}}pd\textbf{S}}{m^{vp}_{i}},
\end{equation}
where $ \textbf{a}^{vp} $ denotes the acceleration, $ \textbf{v}^{vp} $ the velocity vector, $ p $ the pressure, $ \rho $ the density, $ \Omega_{i} $ the region corresponding to Voronoi particle $ i $ and $ \partial\Omega_{i} $ the Voronoi cell surface. The pressure of the Voronoi particle is defined as

\begin{equation}
\label{eq:VP_pressure}
p^{vp}_{i}=\dfrac{m^{vp}_{i}}{m^{vp}_{tg,i}},
\end{equation}
where $ m^{vp}_{tg,i} $ is the target mass. The pressure at the surface between two neighboring cells is computed by second-order approximation $ p^{vp}_{ij}=(p^{vp}_{i}+p^{vp}_{j})/2 $. The scale $ h^{vp}_{i} $ for Voronoi particle $ i $ is defined as the average distances from all neighboring particles.

In each interation substep the Voronoi particles first are moved according to the VP method by
\begin{equation}
\label{eq:VP_update_1}
\textbf{x}_{i}^{vp,*}=\textbf{x}_{i}^{vp,n}+\alpha\dfrac{1}{2}\textbf{a}_{i}^{vp,n} \tau_{1}^{2},
\end{equation}
and then updated following CVT construction as
\begin{equation}
\label{eq:VP_update_2}
\textbf{x}_{i}^{vp,n+1}=\textbf{x}_{i}^{vp,*}+(1-\alpha) \tau_{2}(\textbf{z}_{i}^{vp,n}-\textbf{x}_{i}^{vp,*}),
\end{equation}
where $ \tau_{1} $ and $ \tau_{2} $ are pseudo timestep sizes. The relaxation parameter $ \alpha $ is set as 0.8.

\subsection{Imbalance monitoring strategy}
\label{S:2_1}

To facilitate dynamic load balancing in the current framework, we develop an imbalance monitoring system. Two criteria are constructed to indicate the imbalance of a computation: (1) imbalance caused by the load change of SPH particles in $ \Omega_{i} $, defined as $ m^{vp}_{i} $; (2) imbalance due to the change of communication, i.e. the number of ghost buffer particles constructed in $ \Omega_{i} $, defined as $ mc^{vp}_i $.

The computational load for each SPH particle $ l^{sp}_{j,i} $ is estimated by considering two key operations, the \textit{neighbor search} (\textit{NS}) and the \textit{calculation of inter-particle forces} (\textit{CF}),
\begin{equation}
\label{eq:mass_sph_particle}
l^{sp}_{j,i}=\epsilon \overline{l}^{sp}_{j,i,FNS} + (1-\epsilon) \overline{l}^{sp}_{j,i,SE},
\end{equation}
where $ \overline{l}^{sp}_{\lbrace\cdot\rbrace} $ denotes the normalized computational load. The adaptive weight $ \epsilon $ is obtained by
\begin{equation}
\label{eq:partition_weight}
\epsilon = \dfrac{\Delta t_{NS}}{\Delta t_{NS}+\Delta t_{CF}},
\end{equation}
where $ \Delta t_{\lbrace\cdot\rbrace} $ denotes the net runtime elapsed for different subroutines since the last load-balance estimate. The total computational-load for $ \Omega_{i} $ is calculated by
\begin{equation}
\label{eq:VP_mass}
m^{vp}_{i}=\int_{\Omega_{i}}\rho(\textbf{x})d\sigma=\sum_{j=0}^{N_{i}-1}l^{sp}_{j,i},
\end{equation}
where $ N_{i} $ is the number of SPH particles included in the Voronoi cell $ i $.

To combine the two criteria, an imbalance monitoring tag $ R $ is defined in Eq. \ref{eq:partition_criteria}. The CVP method is triggered when $ R = 1 $.

\begin{equation}
\label{eq:partition_criteria}
R=
\begin{cases}
1& \text{if $ E_{mc^{vp},max} > e_{mc^{vp},max} $ or $ E_{m^{vp},max}>e_{m^{vp},max} $}\\
0& \text{if else}
\end{cases}
,
\end{equation}
\begin{equation}
\label{eq:error_max_communication}
E_{mc^{vp},max} = max(E_{mc^{vp},0},...,E_{mc^{vp},k-1}),
\end{equation}
\begin{equation}
\label{eq:error_max_computation}
E_{m^{vp},max} = max(E_{m^{vp},0},...,E_{m^{vp},k-1}),
\end{equation}
where $ E_{mc^{vp},i} = \dfrac{mc^{vp}_{i}-mc_{0,i}^{vp}}{mc_{0,i}^{vp}} $ and $ E_{m^{vp},i} = \dfrac{m^{vp}_{i}-m_{0,i}^{vp}}{m_{0,i}^{vp}} $, with $ i = 0,...,k-1 $ are local errors in each sub-domain. $ mc_{0,i}^{vp} $ and $ m_{0,i}^{vp} $ are initial values set after each partitioning. $ e_{mc^{vp},max} $ and $ e_{m^{vp},max} $ are user defined error tolerance respectively. In current framework, we set $ e_{mc^{vp},max} = e_{m^{vp},max} = 0.1 $.

\section{Lagrangian Inertial CVP (LICVP) method}
\label{S:3}

In this section, the detailed formulation of the LICVP method is developed. First, a three-step time integration scheme is proposed by introducing a background velocity for advecting Voronoi particles. The Inertial CVP is described in subsection \ref{S:3_2}. A splitting operator and an adaptive filter is defined to optimize and select partitioning strategy dynamically.

\begin{figure}[h!]
\centering
\includegraphics[width=0.8\textwidth]{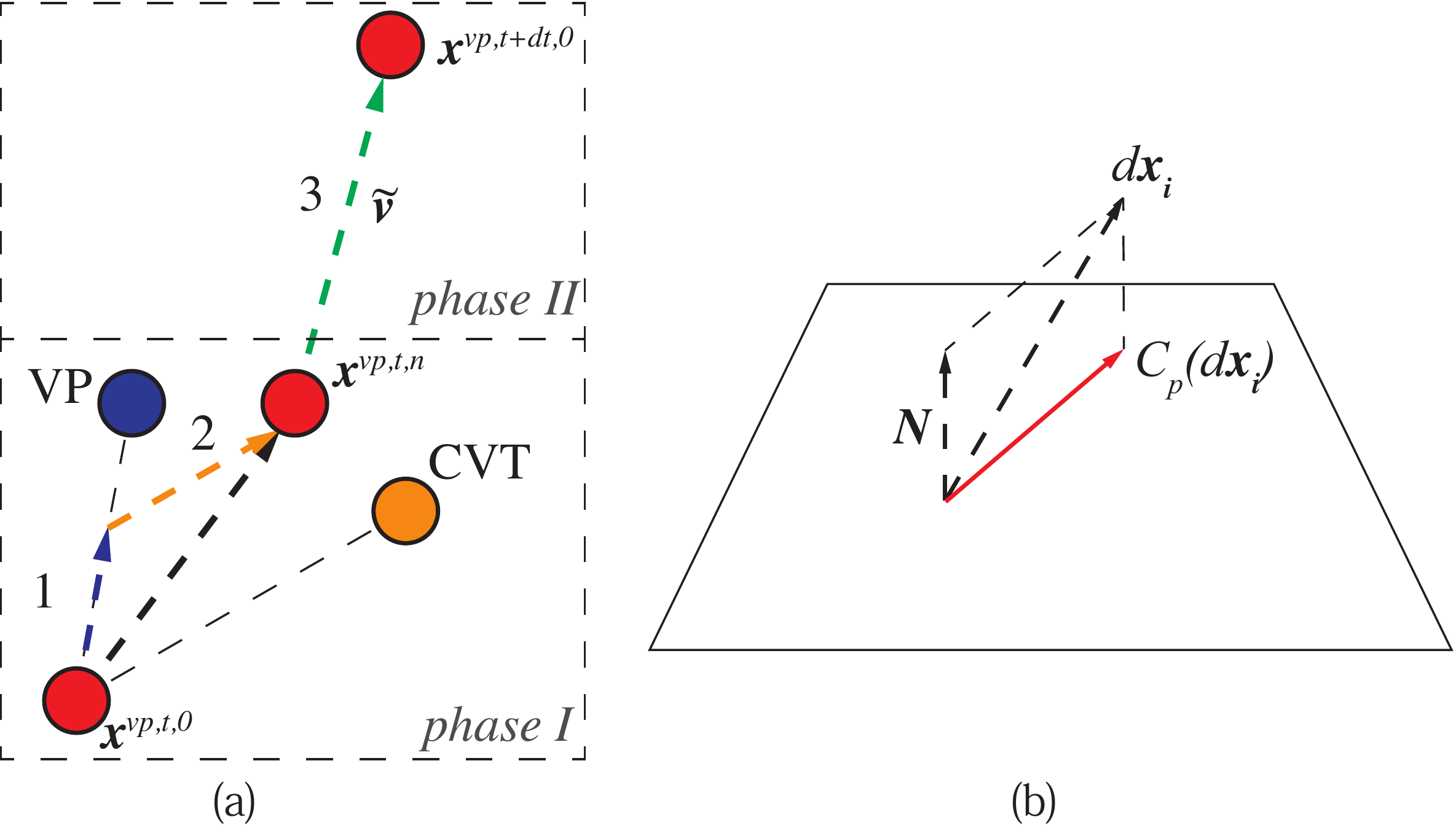}
\caption{(a) Demonstration of three-step time integration scheme. (b) Demonstration of the operator $ C_p(\textbf{A}) $ if partitioning on the plane $ \perp \textbf{N} $.}
\label{Fig:S3_lagrangian_cvp}
\end{figure}

\subsection{Background velocity for Voronoi particles}
\label{S:3_1}

We first define two phases in dynamic load balancing using the CVP method (illustrated in Fig. \ref{Fig:S3_lagrangian_cvp} (a)). The first is the partitioning phase and the second is the load imbalance monitoring phase. In the first phase, the partitioning subroutine is triggered. The initial position of Voronoi particle $ i $ at time $ t $, denoted as $ \textbf{x}^{vp,t,0}_{i} $, is evolved using a two-step integration scheme given by Eq. \ref{eq:VP_update_1} and \ref{eq:VP_update_2}. The partition operation is terminated when equilibrium is achieved after $ n $ pseudo time-steps. The final position of Voronoi particle $ i $ is $ \textbf{x}^{vp,t,n}_{i} $, and the computational domain is repartitioned accordingly. In the second phase, the imbalance monitoring system is launched. The imbalance errors $ E_{mc^{vp},i} $ and $ E_{m^{vp},i} $ are examined during the computation. The error accumulates until the threshold is reached, i.e. $ R = 1 $, then phase 1 is activated again. The final position of Voronoi particle $ i $ in phase 2 is marked as $ \textbf{x}^{vp,t+\Delta t,0}_{i} $.

For phase I, we develop a two-step time integration scheme to achieve both load balance and communication reduction targets. To increase the data reuse in rebalancing, we propose a third step to advect Voronoi particles in phase II (see Fig. \ref{Fig:S3_lagrangian_cvp} (a)). The third time integration step can be defined as
\begin{equation}
\label{eq:VP_position}
\textbf{x}^{vp,t + \Delta t,0}_{i}=\textbf{x}^{vp,t,n}_{i} + \widetilde{\textbf{v}}^{vp}_{i}\cdot\Delta t,
\end{equation}
where $ \Delta t $ is the timestep size identical to the underlined dynamic system, e.g. the SPH flow model. $ \widetilde{\textbf{v}}^{vp}_{i} $ is a background velocity for Voronoi particle $ i $, which can be given arbitrarily. However, to preserve the incremental property, we suggest that $ \widetilde{\textbf{v}}^{vp}_{i} $ is correlated to the motion of $ \Omega_{i} $.

Several options for calculating $ \widetilde{\textbf{v}}^{vp}_{i} $ can be considered. The first is to simply set the background velocity of Voronoi particle $ i $ identical to the mean fluid velocity within the current subdomain.
\begin{equation}
\label{eq:VP_backgroud_velocity_2}
\widetilde{\textbf{v}}^{vp}_{i}=\dfrac{\sum_{j=0}^{N_{i}-1}\textbf{v}^{sp}_{j,i}}{N_{i}},
\end{equation}
where $ \textbf{v}^{sp}_{j,i} $ is the velocity of SPH particle $ j $ in subdomain $ \Omega_{i} $.

Another option is to consider the influence of computational load distribution. The mass center of $ \Omega_{i} $ is given as
\begin{equation}
\label{eq:VP_mass_center}
\textbf{z}^{vp}_{i}=\dfrac{\int_{\Omega_{i}}\rho(\textbf{x})\textbf{x}d\sigma}{\int_{\Omega_{i}}\rho(\textbf{x})d\sigma}=\dfrac{\sum_{j=0}^{N_{i}-1}l^{sp}_{j,i}\textbf{x}^{sp}_{j,i}}{m^{vp}_{i}},
\end{equation}
where $ d\sigma $ denotes the volume differential, and $ \textbf{x}^{sp}_{j,i} $ the coordinates of specific SPH particle $ j $ in $ \Omega_{i} $. Then $ \widetilde{\textbf{v}}^{vp}_{i} $ can be defined by the time differential of mass center,
\begin{equation}
\label{eq:VP_backgroud_velocity_1}
\widetilde{\textbf{v}}^{vp}_{i}=\dfrac{d\textbf{z}^{vp}_{i}}{dt}=\dfrac{\sum_{j=0}^{N_{i}-1}l^{sp}_{j,i}\textbf{v}^{sp}_{j,i}}{m^{vp}_{i}}.
\end{equation}
In practice, instead of updating the Voronoi particles according to the velocity of the mass center of every timestep, we simply set the mass center as the position of Voronoi particles before repartitioning.

We consider the aforementioned two choices in this paper. The performance and detailed comparison will be given in section \ref{S:5}.

\subsection{Inertial CVP}
\label{S:3_2}

In order to constrain the motion of Voronoi particle, we define a splitting operator, which operates on the time-integration scheme in phase I. We can rewrite Eq. \ref{eq:VP_update_1} and \ref{eq:VP_update_2} as
\begin{equation}
\label{eq:ICVP_update_1_step}
\textbf{x}^{vp,n+1}_{i}=\textbf{x}^{vp,n}_{i} + \Delta \textbf{x}^p_{i},
\end{equation}
and 
\begin{equation}
\label{eq:ICVP_operater}
\Delta \textbf{x}^p_{i} = C_p(\Delta \textbf{x}_{i}),
\end{equation}
where $ \Delta \textbf{x}^p_{i} $ is the resulting displacement. The splitting operator $ C_p $ functions via manipulating $ \Delta \textbf{x} $, where $\Delta \textbf{x}_{i} = \alpha\dfrac{1}{2}\textbf{a}_{i}^{vp,n}\bigtriangleup \tau_{1}^{2} + (1-\alpha)\bigtriangleup \tau_{2}(\textbf{z}_{i}^{vp,n}-\textbf{x}_{i}^{vp,*})$.

For a general input vector $ \textbf{A} $ and a given direction of constraint, $ C_p $ returns a resultant vector defined as
\begin{equation}
\label{eq:ICVP_operater_defination}
C_p(\textbf{A})=
\begin{cases}
\textbf{A}_\parallel, \text{where} \textbf{A}_\parallel \parallel \textbf{N} & \text{if constrained along $ \textbf{N} $}\\
\textbf{A}_\perp, \text{where} \textbf{A}_\perp \perp \textbf{N}& \text{if constrained on the plane $ \perp \textbf{N} $}\\
\textbf{A}& \text{if no constraint}
\end{cases}
,
\end{equation}
where $ \textbf{A}_\parallel $ and $ \textbf{A}_\perp $ denote the vector parallel and perpendicular to $ \textbf{N} $, respectively. $ \textbf{N} $ is the input that defines the direction of constraint. As illustrated in Fig. \ref{Fig:S3_lagrangian_cvp} (b), the particle motion is constrained on a plane where $ \textbf{N} $ denotes its normal direction. In this case, $ C_p(\textbf{A}) $ returns $ \textbf{A}_\perp $ that is the projection of $ \textbf{A} $ into the plane.

To calculate $ \textbf{N} $, we need to characterize the computational load distribution. In the RIB method \cite{simon1991partitioning}, the inertia matrix is utilized to calculate the principle inertia axis \cite{hendrickson2000dynamic}. Inspired by the RIB method, we can find the direction of constraint by the same means.

The global mass center of simulation is calculated by
\begin{equation}
\label{eq:ICVP_global_mass_center}
\overline{\textbf{Z}}=\dfrac{\int_{\Omega}\rho(\textbf{x})\textbf{x}d\sigma}{\int_{\Omega}\rho(\textbf{x})d\sigma}=\dfrac{\sum_{i=0}^{N-1}m^{vp}_{i}\textbf{z}^{vp}_{i}}{\sum_{i=0}^{N-1}m^{vp}_{i}}.
\end{equation}
Then the inertial matrix is obtained as
\begin{equation}
\label{eq:ICVP_J_matrix}
\textbf{J}=\sum_{k=0}^{n-1}l^{sp}_{k}(\textbf{x}^{sp}_{k}-\overline{\textbf{Z}})(\textbf{x}^{sp}_{k}-\overline{\textbf{Z}})^{T},
\end{equation}
where $ n $ denotes the total number of SPH particles simulated. The eigenvalue $ \lambda $ (Eq. \ref{eq:ICVP_eigenvalue}) and eigenvector $ \xi $ (Eq. \ref{eq:ICVP_eigenvector}) of $ \textbf{J} $ indicate the value and direction of principle inertia, respectively. The eigenvalues are sorted in increasing order.

\begin{equation}
\label{eq:ICVP_eigenvalue}
\lambda=\lbrace \lambda_1, \lambda_2, \lambda_3 \rbrace,
\end{equation}
\begin{equation}
\label{eq:ICVP_eigenvector}
\xi=\lbrace \xi_{\lambda_1}, \xi_{\lambda_2}, \xi_{\lambda_3} \rbrace.
\end{equation}

The direction of the constraint can be calculated accordingly. We propose an inertial-based adaptive filter defined as
\begin{equation}
\label{eq:ICVP_project_vector}
\textbf{N}=
\begin{cases}
\xi_{\lambda_{1}} & \text{if $ \overline{\lambda_{3}} > \lambda_{max}$, i.e. constrain along \textbf{N}}\\
\xi_{\lambda_{3}} & \text{if $ \overline{\lambda_{1}} < \lambda_{min}$, and $\overline{\lambda_{2}} > \lambda_{min} - \overline{\lambda_{1}} $, i.e. constrain on the plane $ \perp \textbf{N} $}
\end{cases}
,
\end{equation}
where $ \overline{\lambda_i} = \dfrac{\lambda_i}{\sum_{i=0}^{2}\lambda_i} $. $ \lambda_{max} $ and $ \lambda_{min} $ are user defined thresholds. The selection procedure is to compare the normalized eigenvalues with predefined thresholds, and determine an appropriate strategy. The filter is adaptive as well, which allows for dynamic optimization of the partitioning strategy via choosing different constraints. For instance, if $ \overline{\lambda_{3}} > \lambda_{max} $, the load along the first principle inertial axis, i.e. $ \xi_{\lambda_{1}} $, is considerably larger than the combination of the other two axis. Thus, the degrees of freedom in $ \xi_{\lambda_{2}} $ and $ \xi_{\lambda_{2}} $ are confined, and only motion along $ \xi_{\lambda_{1}} $ is allowed. During the computation, if the other condition is satisfied, the strategy will be altered accordingly. If both conditions are not satisfied, the original CVP method is restored.

\subsection{Intermediate conclusions}
\label{S:3_3}

In section \ref{S:3_1} and \ref{S:3_2}, we propose two extensions of CVP method. Both algorithms are mutually independent and compatible since they function at different phases of the computation, thus they can be integrated into a single framework. The objective of the LICVP method is to optimize the incremental property in rebalancing without violating the other constrains in static partitioning. In general, the Lagrangian property improves the data reuse when SPH particles move following the fluid field, while the Inertial CVP guarantees that partitioning is aware of the global load distribution and insensitive of the load variation along directions of minimum interest. LICVP methods can restore to original CVP method under certain conditions, and the implementation requires trivial modification. Moreover, The additional cost due to the extension is minimum since updating Voronoi particles is localized within sub-domains and the global inertial matrix is only updated once the repartitioning is triggered.

\section{Numerical algorithms}
\label{S:4}

\subsection{Partitioning boundary conditions}
\label{S:4_1}

\begin{figure}[h!]
\centering
\includegraphics[width=0.8\textwidth]{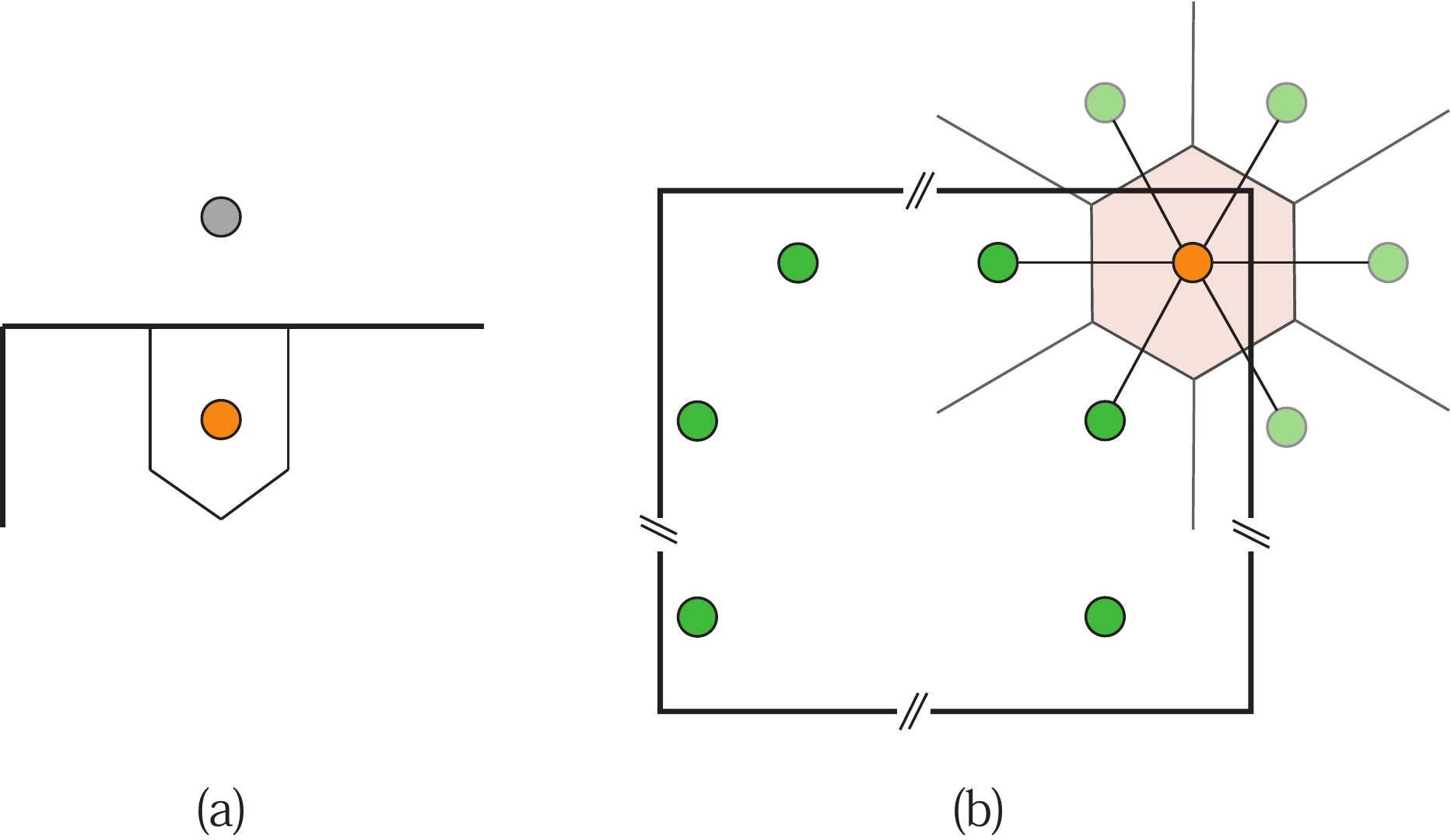}
\caption{Sketch of (a) symmetric and (b) periodic boundary conditions. The orange particles indicate the boundary Voronoi particles; the grey particle is the mirrored particles constructed to enforce symmetric boundary condition; the solid green particles are neighbors of the current (orange) Voronoi particle in a periodic box, and the translucent particles are mapped neighbor particles.}
\label{Fig:S4_bc}
\end{figure}

Symmetric and periodic boundary conditions can be applied on the partitioning domain (see Fig. \ref{Fig:S4_bc}). The ghost particle method is employed to enforce both boundary conditions.

Ghost particles for symmetric boundary conditions (SBC) are constructed following \cite{fu2017physics}. SBC is employed when the computational domain is symmetric itself or is defined as open boundary.

Periodic boundary conditions (PBC) are enforced in a similar way. In the first step, Voronoi particles that have neighbors on the other side of the periodic boundary are identified, and marked as boundary particles. In the second step, all boundary particles are mapped by adding/subtracting the domain length in the periodic direction, and the Voronoi diagram is constructed again including ghost particles. All physical variables of the ghost particles are identical to the mapped boundary particles. When the background velocity is introduced, PBC is enforced for phase II as well, which facilitates Voronoi particles for better tracing the local sub-domains. The benefit of PBC is twofold: 1) SPH particles that transported across the boundary during simulation cost no extra inter-processor data relocation; 2) the rebalancing is calculated allowing Voronoi particles to move across the boundary, thus data reuse is better preserved.

\subsection{Flowchart}
\label{S:4_2}

\begin{algorithm}[h!]\footnotesize
\caption{Flowchart of LICVP method for one timestep}
\label{alg:LICVP}
\begin{algorithmic}[1]

\If{current timestep count is divisible by 20} 
  \Comment {\textit{the imbalance monitoring system is activated every 20 timesteps}}
  \State calculate imbalance error $ E_{mc^{vp},i} $ (Eq. \ref{eq:error_max_communication}) and $ E_{m^{vp},i} $ (Eq. \ref{eq:error_max_computation});
  \State Check whether repartitioning is required, i.e. $ R=1 $ (Eq. \ref{eq:partition_criteria});
  \If{$ R=1 $} \Comment {\textit{the system will be rebalanced}}
    \State calculate the inertial matrix $ J $ (Eq. \ref{eq:ICVP_J_matrix}), and find its eigenvalue and eigenvector (Eq. \ref{eq:ICVP_eigenvalue} and \ref{eq:ICVP_eigenvector});
    \State calculate direction of constraint, i.e. $ \textbf{N} $ (Eq. \ref{eq:ICVP_project_vector}):
    \While{the partitioning error is not converged}
      \State construct Voronoi diagram and solve the model equation Eq. \ref{eq:VP_pressure};
      \State calculate the resulting displacement $ \Delta \textbf{x}^p_{i} $ according to Eq. \ref{eq:ICVP_operater} and Eq. \ref{eq:ICVP_operater_defination};
      \State update Voronoi particles;
      \State check partitioning error;
    \EndWhile
    \State migrate SPH particles to target processor according to the new partitioning result;
  \EndIf
  \State update data structure and construct ghost buffer particles;
  \State solve SPH governing equations;
  \State calculate the background velocity $ \widetilde{\textbf{v}}^{vp}_{i} $ for transporting Voronoi particles using Eq. \ref{eq:VP_backgroud_velocity_2} or \ref{eq:VP_backgroud_velocity_1};
  \State update the position of Voronoi particles according to Eq. \ref{eq:VP_position};
\EndIf
\end{algorithmic}
\end{algorithm}

The detailed algorithm for LICVP method is summerized in Alg. \ref{alg:LICVP}. For simplicity, we only list the flowchart for one simulation timestep. Other correlative algorithms, e.g. CVP method, data structure, communication strategy, etc., are not elaborated here. One can refer to our previous work \cite{fu2017physics}\cite{linfu_fast_neighbor}\cite{zhe_ji_new} for a comprehensive overview.

\section{Numerical validation}
\label{S:5}

In this section, 5 test cases are considered to validate the proposed LICVP method. The underlining Lagrangian property of LICVP method is demonstrated via case 1, 2 and 3, where the Inertial CVP method degenerates to the original CVP method. The feasibility and performance of Inertial CVP method is then discussed in case 4 and 5. Case 3 is tested in the mpp2 cluster provided by Leibniz-Rechenzentrum (LRZ), which is constructed by 28-way Haswell-EP nodes with Infiniband FDR14 interconnect. The rest cases are carried out on the same workstation with 12 Intel(R) Xeon(R) CPU E5-2630 v2 cores (64G memory and 2.6GHz) and Scientific Linux system (Release 6.8).

Before moving to the results, we first define two measurements to assess the incremental property and optimization for communication reduction, i.e. averaged communication load ($ S_c $) and averaged particle migration ($ S_m $). $ S_c $ and $ S_m $ are defined as
\begin{equation}
\label{eq:S_c}
S_c=\dfrac{\sum_{i=0}^{N-1}\dfrac{N_{i}^{ghost}}{N_i}}{N},
\end{equation}
\begin{equation}
\label{eq:S_m}
S_m=\dfrac{\sum_{i=0}^{N-1}\dfrac{N_{i}^{migrated}}{N_i}}{N},
\end{equation}
where $N_{i}^{ghost} $ and $ N_{i}^{migrated} $ denotes the number of ghost buffer particles constructed and the number of migrated particles respectively.

\subsection{Case 1}
\label{S:5_1}

\begin{figure}[h!]
\centering
\includegraphics[width=1.0\textwidth]{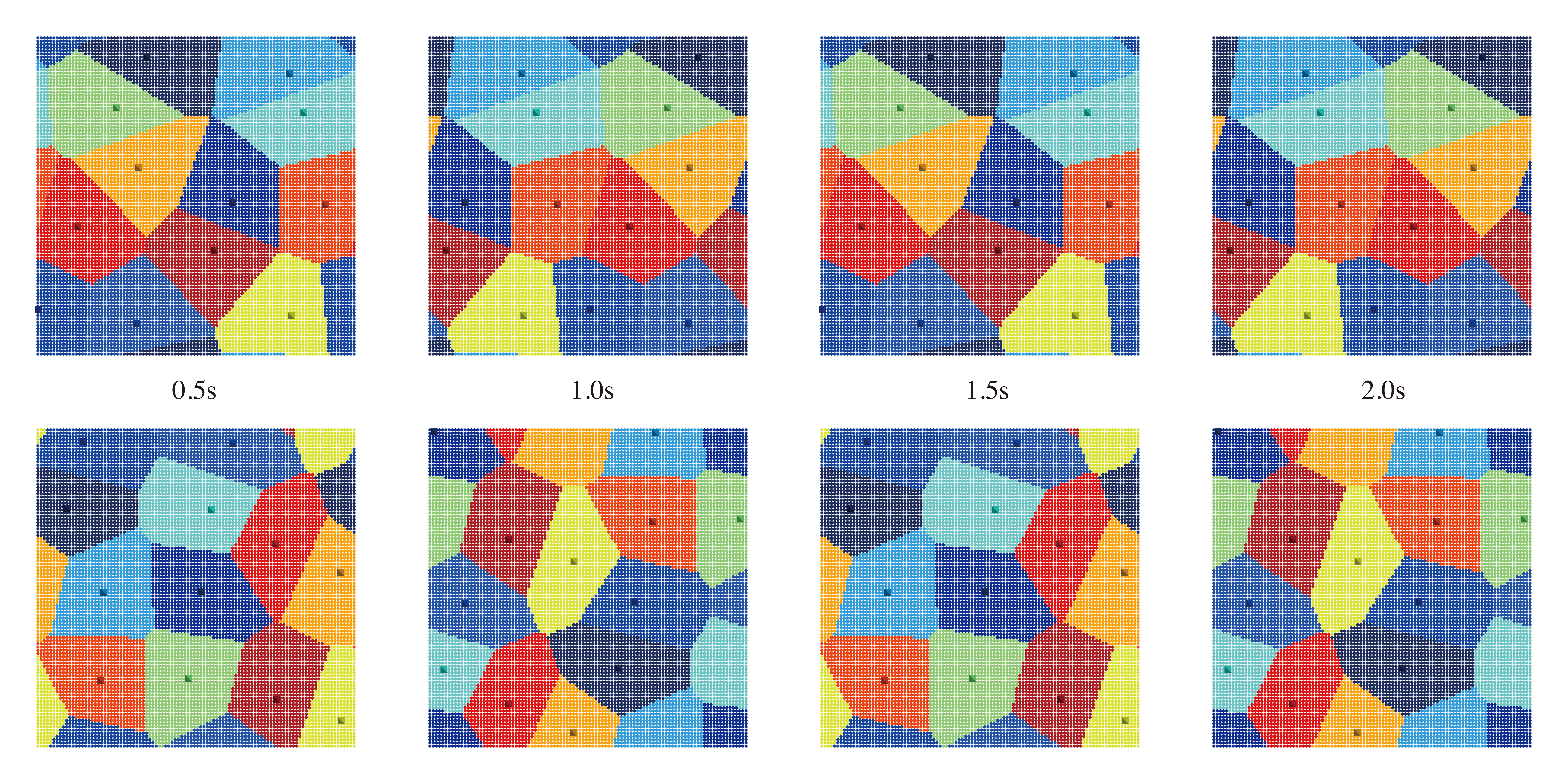}
\caption{Steady advection test: the partitioning result of fluid with horizontal (upper row) and diagonal (bottom row) velocity at four instants. SPH particles are rendered with sub-domain index, and octahedrons are positions of Voronoi particles.}
\label{Fig:S5_steady_flow}
\end{figure}

\begin{figure}[h!]
\centering
\includegraphics[width=0.8\textwidth]{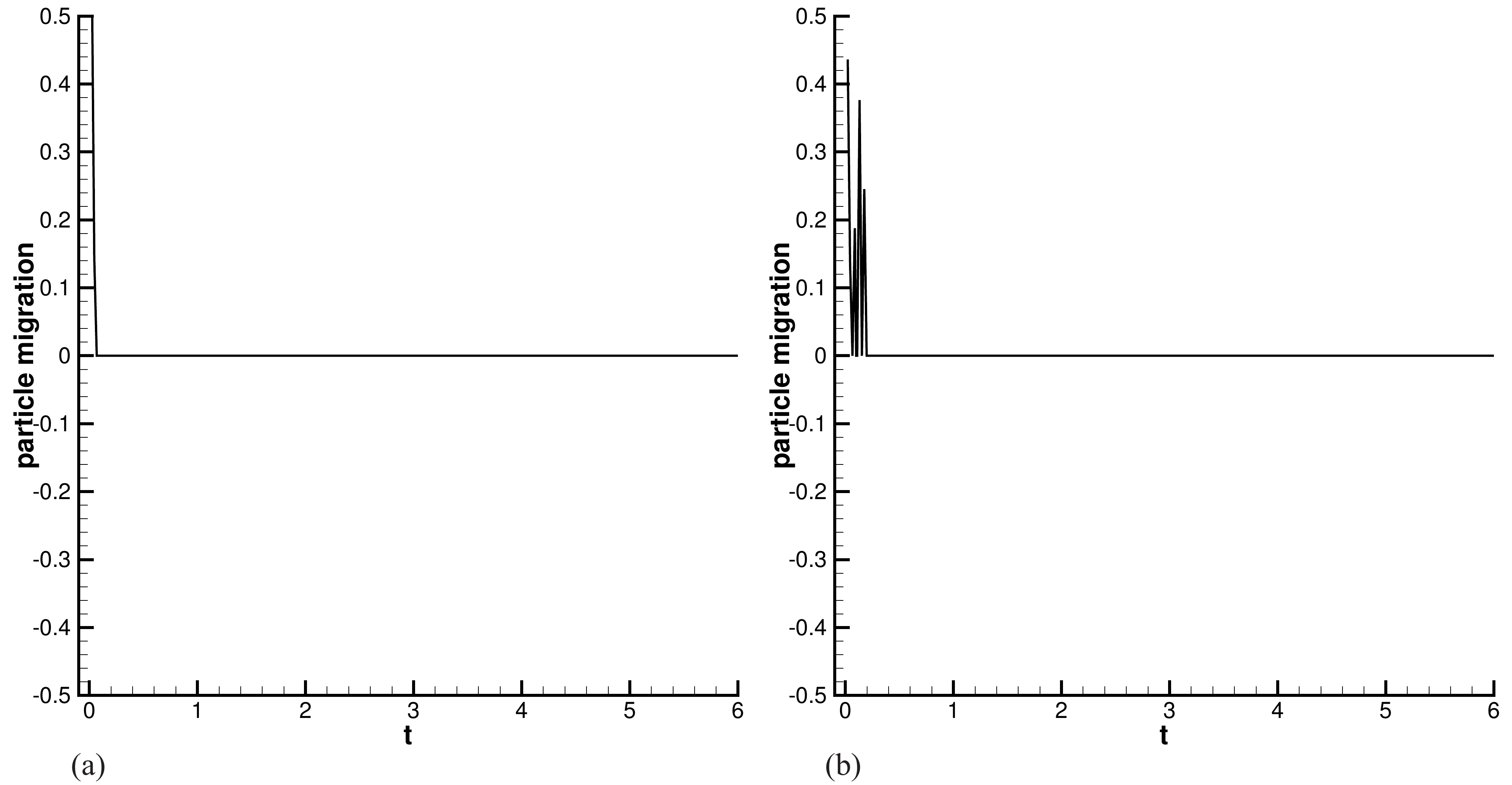}
\caption{Steady advection test: time history of averaged particle migration. (a) fluid with horizontal velocity, (b) fluid with diagonal velocity.}
\label{Fig:S5_steady_flow_migration}
\end{figure}

In the first case, we consider a steady advection test. We initialize a 2D periodic box of unit length. The fluid is assigned with constant density ($ \rho=1000 $), and advected with uniform bulk velocity. A weakly compressible SPH solver is employed following \cite{adami2012generalized}. The number of particles simulated is 10000, and 12 MPI tasks are launched. The partitioning domain is set with equal size of the fluid field and PBC is enforced in both direction.

Since the fluid is advected with constant velocity, the relative position of SPH particles remains identical in entire simulation. With the background velocity, the topology of Voronoi particles should be invariant as well, consequently, after rebalancing, we should obtain the same partitioning diagram every time, and inter-processor particle migration is zero, i.e. data reuse is maximized.

Two bulk velocities, i.e. horizontal ($ \{u,v\}=\{1,0\} $) and diagonal ($ \{u,v\}=\{1,1\} $) velocity, are studied. It is noticed that all the computational load for SPH particle is the same, the mean velocity of the fluid coincides with the velocity of mass center. We rebalance the simulation every 100 timesteps. The partitioning results at four instants are illustrated in Fig. \ref{Fig:S5_steady_flow}. As anticipated, the topology of the partitioning diagram remains unaltered, and repeats every one second. The averaged particle migration (see Fig. \ref{Fig:S5_steady_flow_migration}) is zero during entire simulation, except for a small oscillation at the beginning of the second case.

\subsection{Case 2}
\label{S:5_2}

\begin{figure}[h!]
\centering
\includegraphics[width=1.0\textwidth]{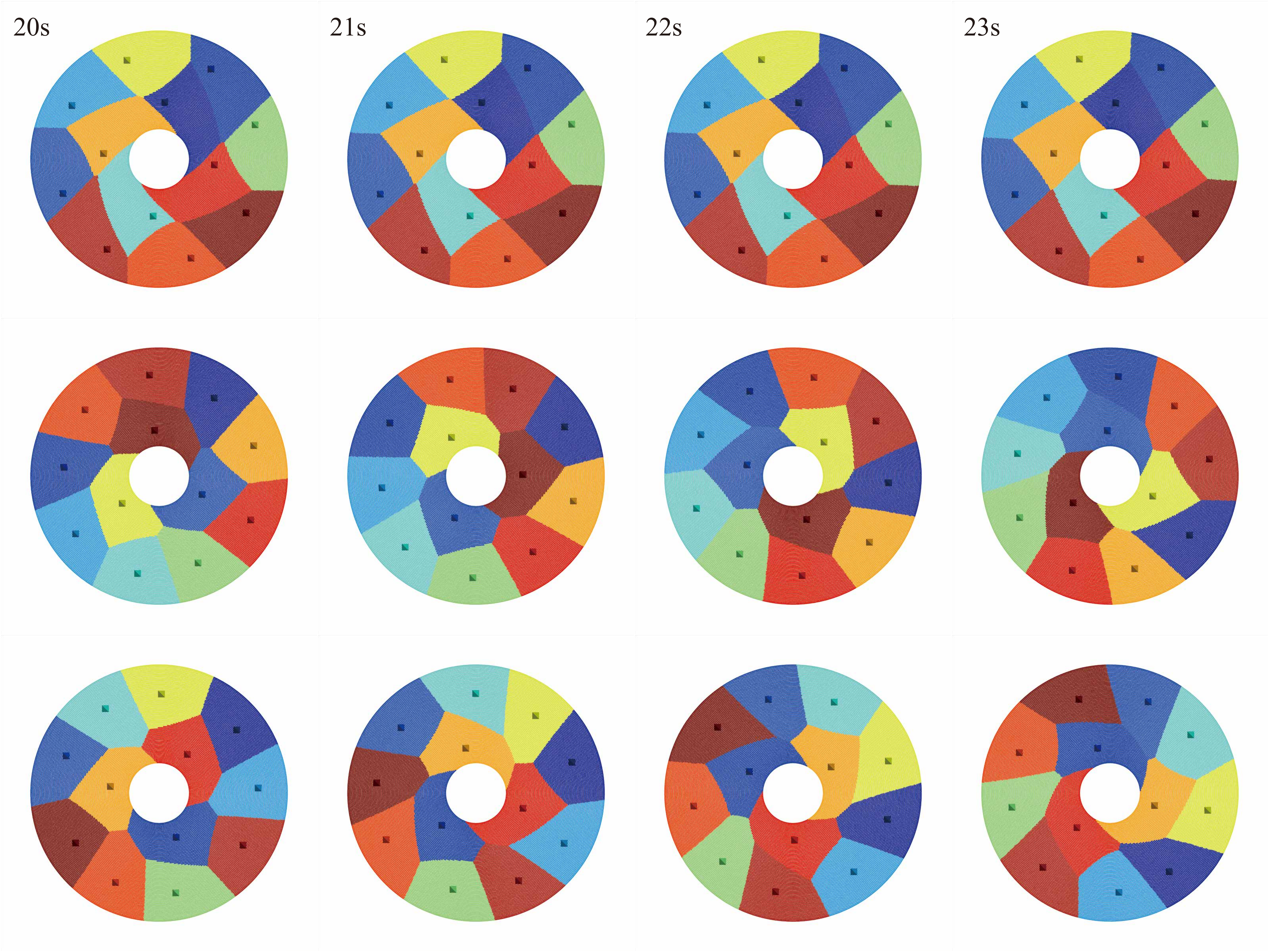}
\caption{2D Keplerian disk problem: comparison between non-moving Voronoi particles (upper row), moving Voronoi particles with velocity of mass center (middle row) and mean fluid velocity (bottom row). Four snapshots are illustrated, i.e. 5s, 6s, 7s and 8s. SPH particles corresponding to different sub-domains are assigned with distinct colors, and the octahedrons denote the Voronoi particles.}
\label{Fig:S5_steady_kepler_ring}
\end{figure}

\begin{figure}[h!]
\centering
\includegraphics[width=0.8\textwidth]{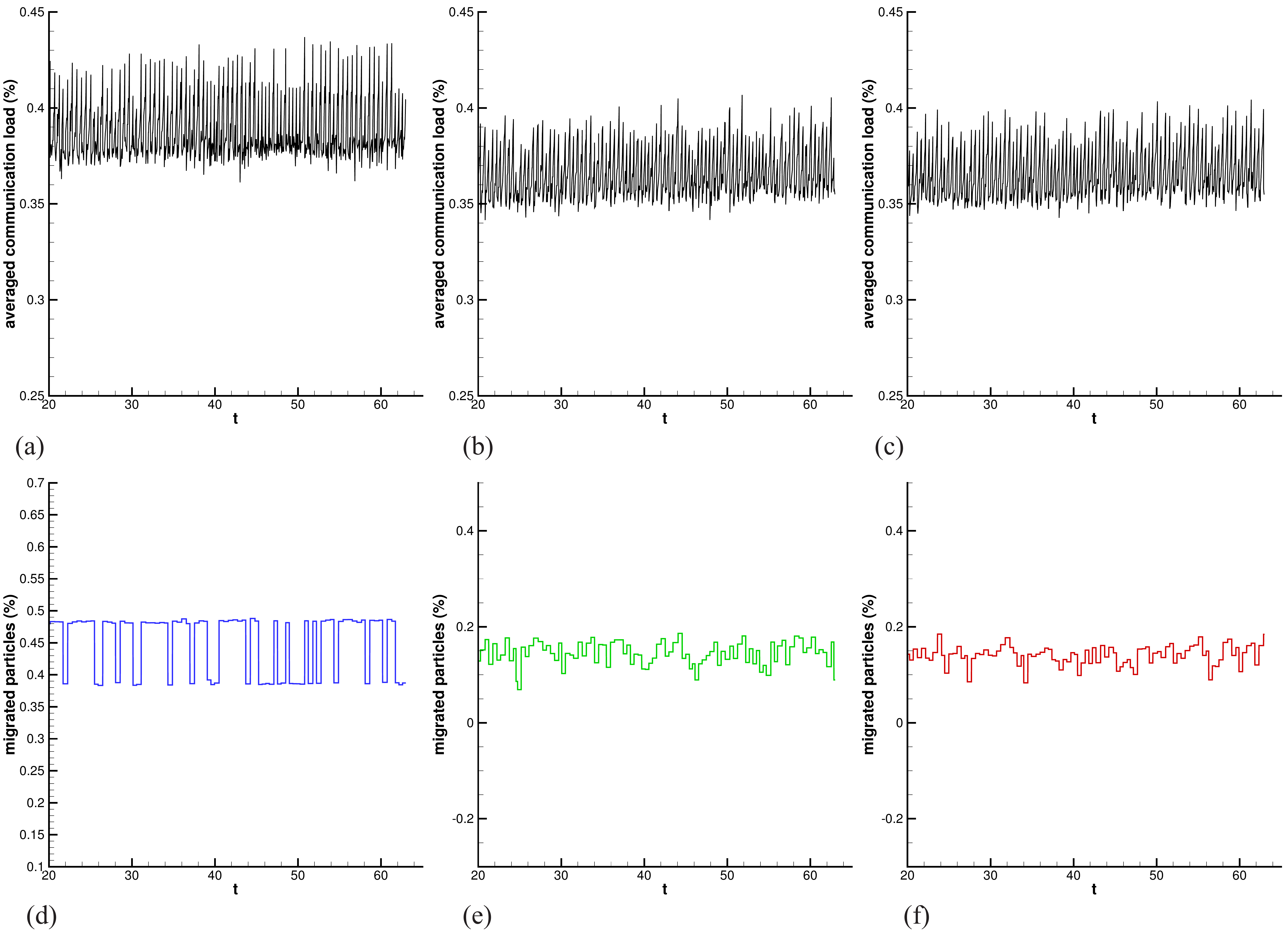}
\caption{2D Keplerian disk problem: The time history of averaged communication load percentage using non-moving Voronoi particles (a), moving Voronoi particles with velocity of mass center (b) and mean fluid velocity (c). The time history of averaged migration percentage using non-moving Voronoi particles (d), moving Voronoi particles with velocity of mass center (e) and mean fluid velocity (f).}
\label{Fig:S5_steady_kepler_ring_curve}
\end{figure}

We consider a 2D cold Keplerian disc problem. It is a typical case in astrophysics, where gas orbits a central point mass subjecting to the equilibrium of gravity, centrifugal force and pressure force \cite{cullen2010inviscid}. We initialize a uniform density disk following \cite{raskin2016examining}. A compressible SPH solver proposed by \cite{price2012smoothed} is employed, where artificial viscosity and conductivity are switched off. To stabilize the flow, a damping term is added to the radial component of particle acceleration \cite{raskin2016examining}. A total number of 47500 particles are simulated with 12 MPI tasks.

Since the flow is shearing, different orbiting velocity will cause the deformation of sub-domains and an increasing of communication load. The proposed two background velocities are chosen to compare with simulation without setting background velocity, i.e. $ \widetilde{\textbf{v}}^{vp}_{i}= \textbf{0}$. The snapshot with regarding to three situations at four instants are illustrated in Fig \ref{Fig:S5_steady_kepler_ring}. The top row gives the result of $ \widetilde{\textbf{v}}^{vp}_{i}= \textbf{0}$. It is observed that the positions of Voronoi particles remain approximately constant after rebalancing and the topology of partitioning result is exactly the same. Conversely, if the Voronoi particles are advected with the background velocity, both results exhibit shifted partitioning sub-domains according to the local flow, and slight topology alteration in a long run. Statistics comparison is manifested in Fig. \ref{Fig:S5_steady_kepler_ring_curve}. All three cases demonstrate that after rebalancing, the communication load decreases instantly, and restores to approximately the same value. Moreover, $ S_c $ exhibits a slightly larger value when the Voronoi particles are not updated comparing to the other two situations. The $ S_m $ for the first case shows a portion of 39\% to 48\% particles are being migrated after each rebalancing. This number drops to approximately 15\% for the other two cases. It can be concluded that with the background velocity, the incremental property of CVP method is improved. The results from two underlying background velocities exhibit slight differences with respect to performance.

\subsection{Case 3}
\label{S:5_3}

\begin{figure}[h!]
\centering
\includegraphics[width=1.0\textwidth]{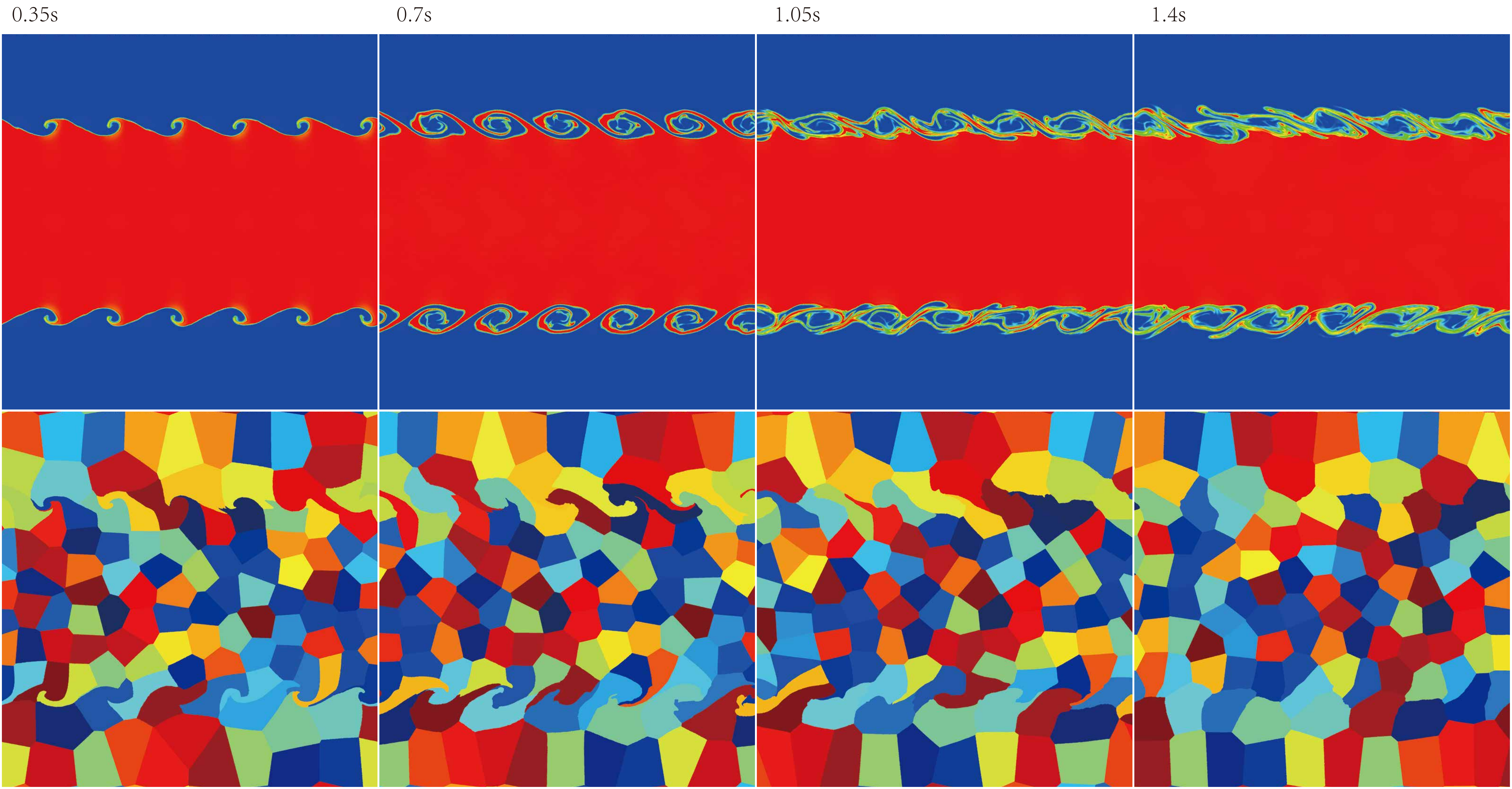}
\caption{2D KH instability: the density field (upper row) and partitioning diagram (bottom row) with respect to four instants.}
\label{Fig:S5_KH_02}
\end{figure}

\begin{figure}[h!]
\centering
\includegraphics[width=0.8\textwidth]{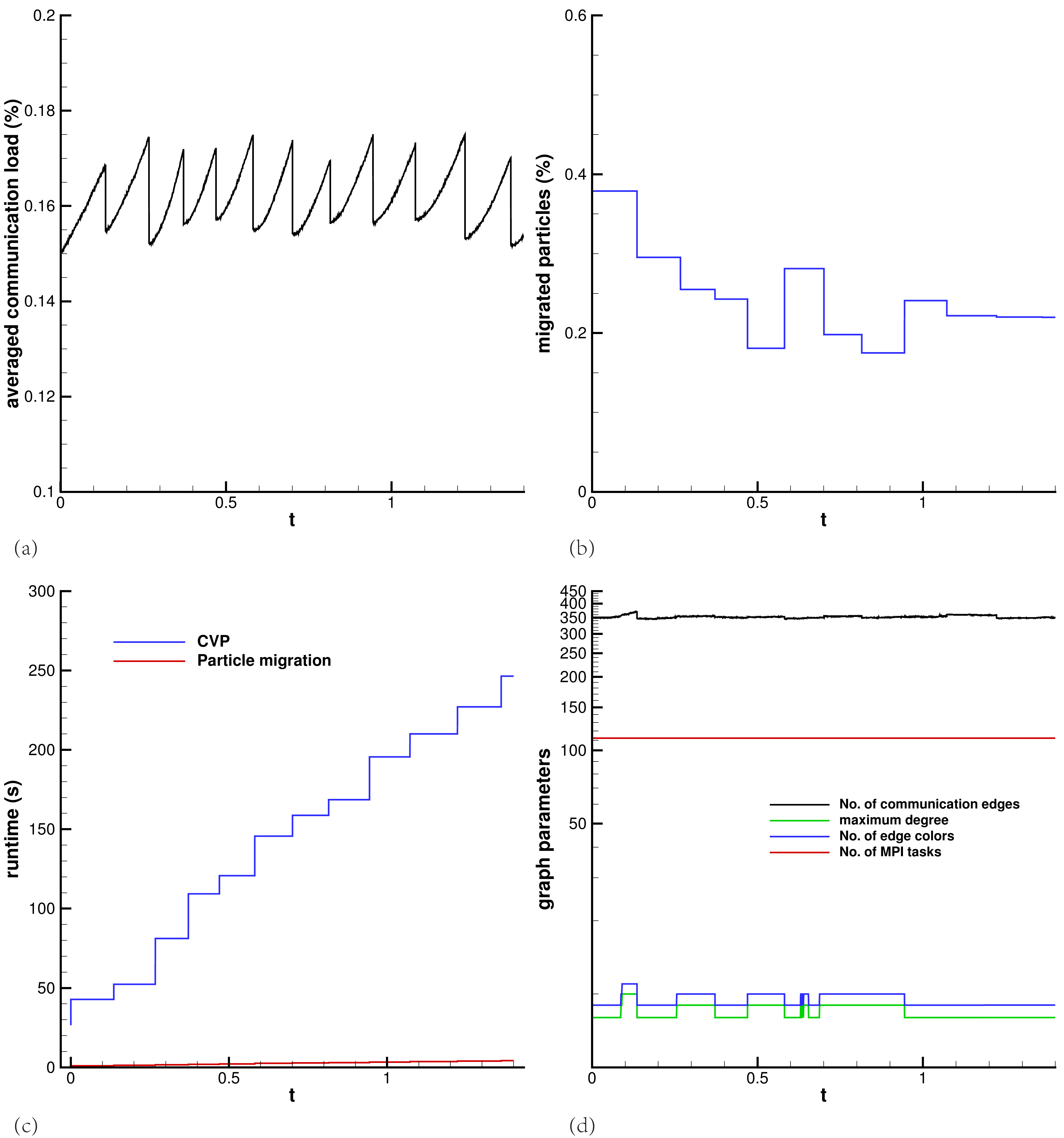}
\caption{2D KH instability: (a) The time history of averaged communication load percentage. (b) the time history of averaged particle migration percentage. (c) runtime of CVP method and particle migration during simulation. (d) time history of graph parameters.}
\label{Fig:S5_KH_01}
\end{figure}

We consider a 2D Kelvin-Helmholtz instability problem following \cite{price2012smoothed}. A Godunov SPH with second order reconstruction solver is employed \cite{inutsuka2002reformulation}\cite{zhe_ji_new}. PBC is enforced at the boarder of partitioning domain. Eq. \ref{eq:VP_backgroud_velocity_2} is utilized to calculate the background velocity. 629,0560 SPH particles are simulated on 112 processors. Each processor has 1 TBB thread.

Fig. \ref{Fig:S5_KH_02} presents the simulation result as well as the partitioning diagram at four instants. The topological layout of sub-domains in smooth region manifests superior similarity, and sub-domains are drifted with the local flow. In regions with discontinuity, due to the development of instability, the partitioning result exhibits topological alteration in a long run. In the simulation duration, i.e. 1.4s, the system is repartitioned 11 times, and $ S_c $ drops immediately after each rebalancing (see Fig. \ref{Fig:S5_KH_01} (a)). The averaged data migration drops from 40\% at the beginning to approximate 25\% after 0.5s (see Fig. \ref{Fig:S5_KH_01} (b)). The runtime for CVP method as well as data migration (illustrated in Fig. \ref{Fig:S5_KH_01} (c)) is negligible comparing to the total simulation time (60559s). The graph parameter, which is introduced in Ref. \cite{zhe_ji_new} and characterizes the sparse communication relationship, demonstrates that the total number of communication relation is bounded and is about 3 times larger than the MPI task number (see Fig. \ref{Fig:S5_KH_01} (d)). The communication finishes within 10 sub-steps during the entire simulation.

\subsection{Case 4}
\label{S:5_4}

\begin{figure}[h!]
\centering
\includegraphics[width=0.8\textwidth]{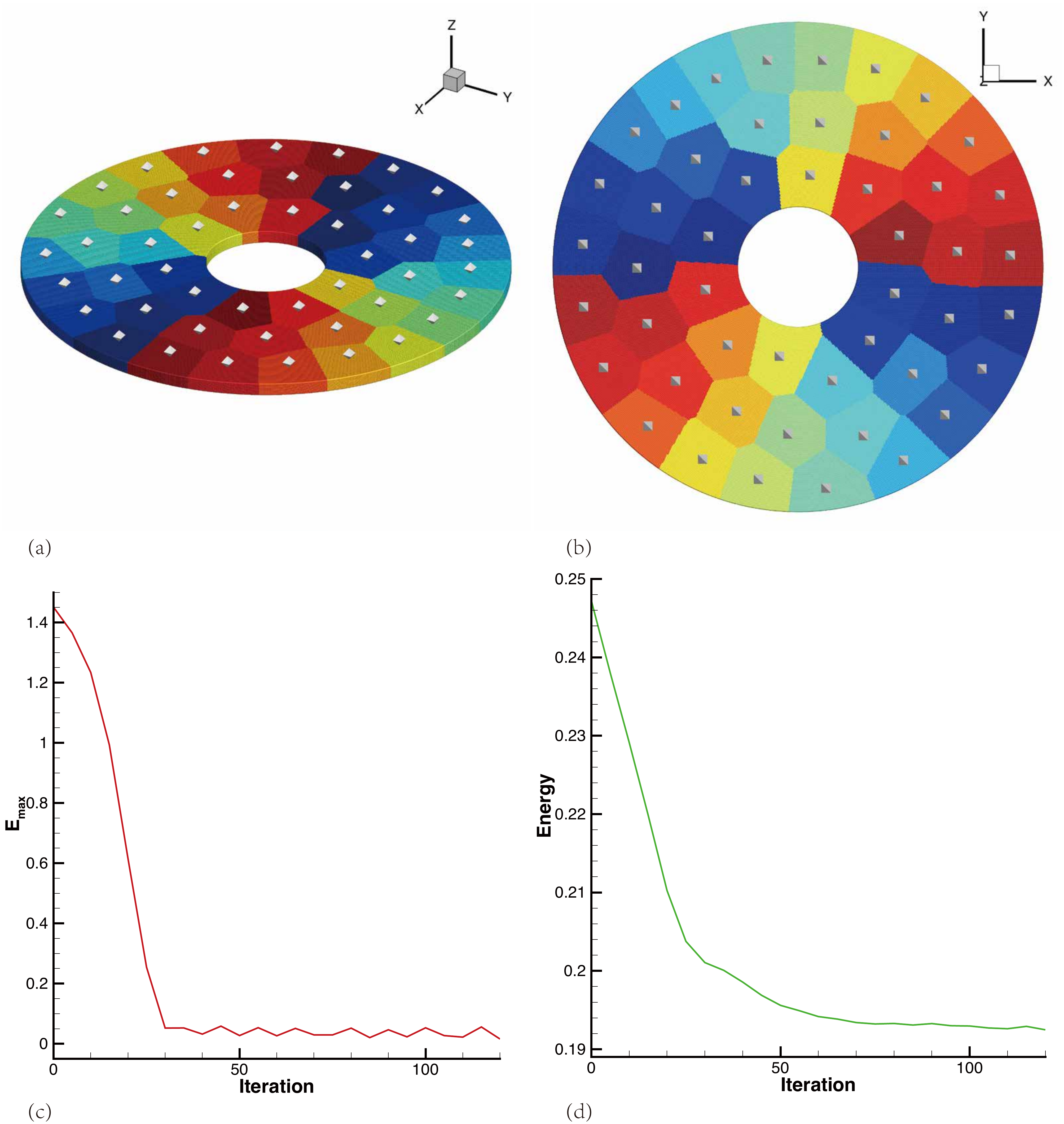}
\caption{3D Keplerian disk problem: the disk is initialized on x-o-y plane. Partitioning result (a)(b). Time history of partitioning error (c) and energy (d).}
\label{Fig:S5_domain_reduction_xy}
\end{figure}

\begin{figure}[h!]
\centering
\includegraphics[width=0.8\textwidth]{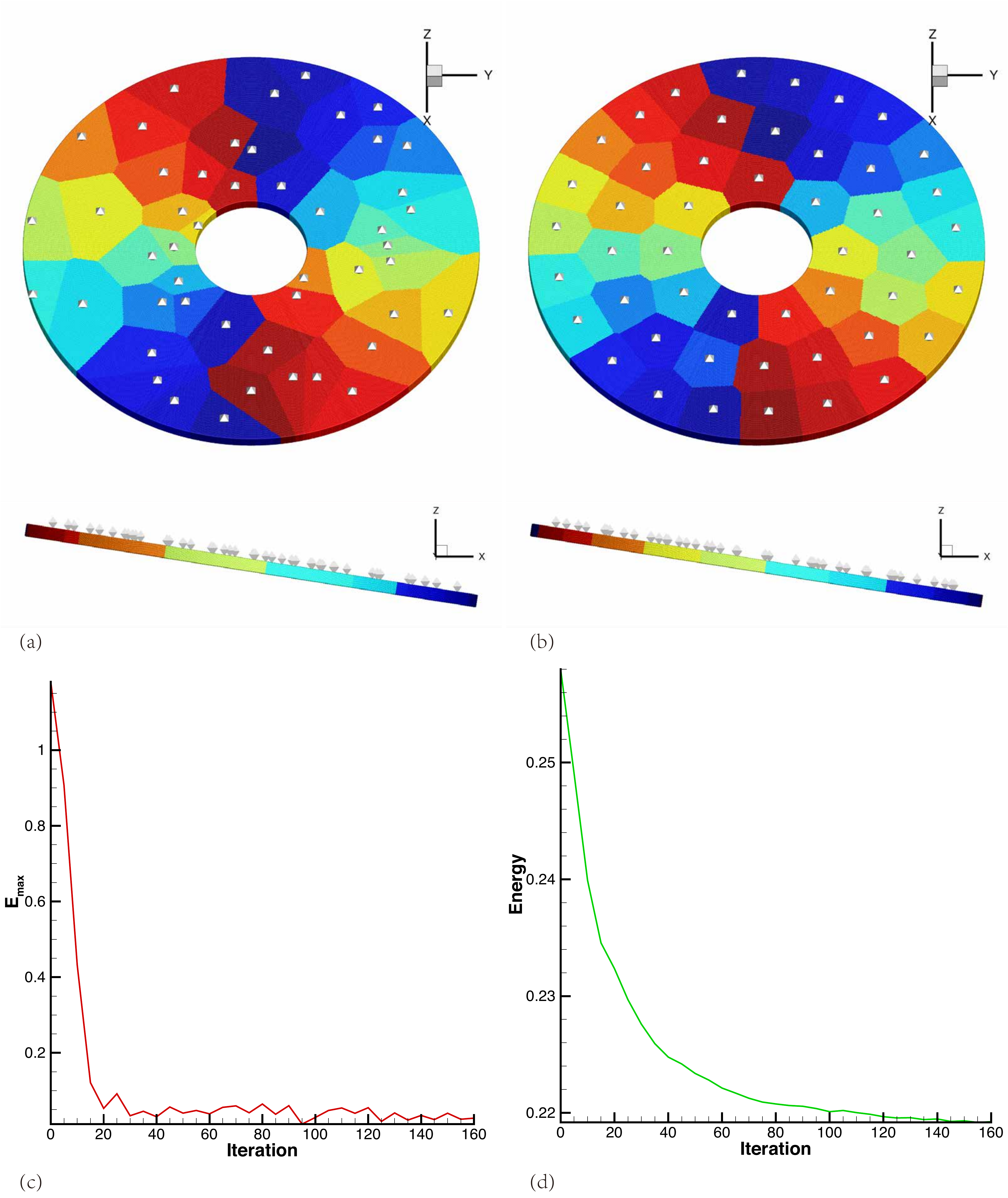}
\caption{3D Keplerian disk problem: the disk is initialized on plane with angle $ \pi/4 $ to x-o-y plane. Initial condition and Voronoi particle distribution (a). Partitioning result (b). Time history of partitioning error (c) and energy (d).}
\label{Fig:S5_domain_reduction_skew}
\end{figure}

The 3D Keplerian disk problem is employed to demonstrate the feasibility of proposed Inertial CVP method. We use the same setup in case 2, and extend the height of the disk to 0.1 in z direction with identical particle pitch. The dynamic evolution of this case gives the same conclusions as in case 2, thus we only focus on the initial partitioning. Two situations are calculated, where the disk is initialized on x-o-y plane and plane with angle $ \pi/4 $ to x-o-y plane respectively. The calculation is carried out with 12 MPI tasks. The Voronoi particles are initialized on disk plane with constant angular separation and a random radius ranging between the outer circle and the inner circle. We set $ \lambda_{max}=0.9 $ and $ \lambda_{min}=0.1 $ in this case.

The result for both cases are plotted in Fig \ref{Fig:S5_domain_reduction_xy} and Fig. \ref{Fig:S5_domain_reduction_skew} respectively. It can be observed that Inertial CVP method captures the load distribution successfully in both cases, where the motion of Voronoi particles is constrained on the disk plane. The partitioning results feature convex, well-shaped sub-domains identical to the original CVP method. The partitioning error converges rapidly. The energy descends monotonically as well, which essentially optimizes the communication volume.

\subsection{Case 5}
\label{S:5_5}

\begin{figure}[h!]
\centering
\includegraphics[width=1.0\textwidth]{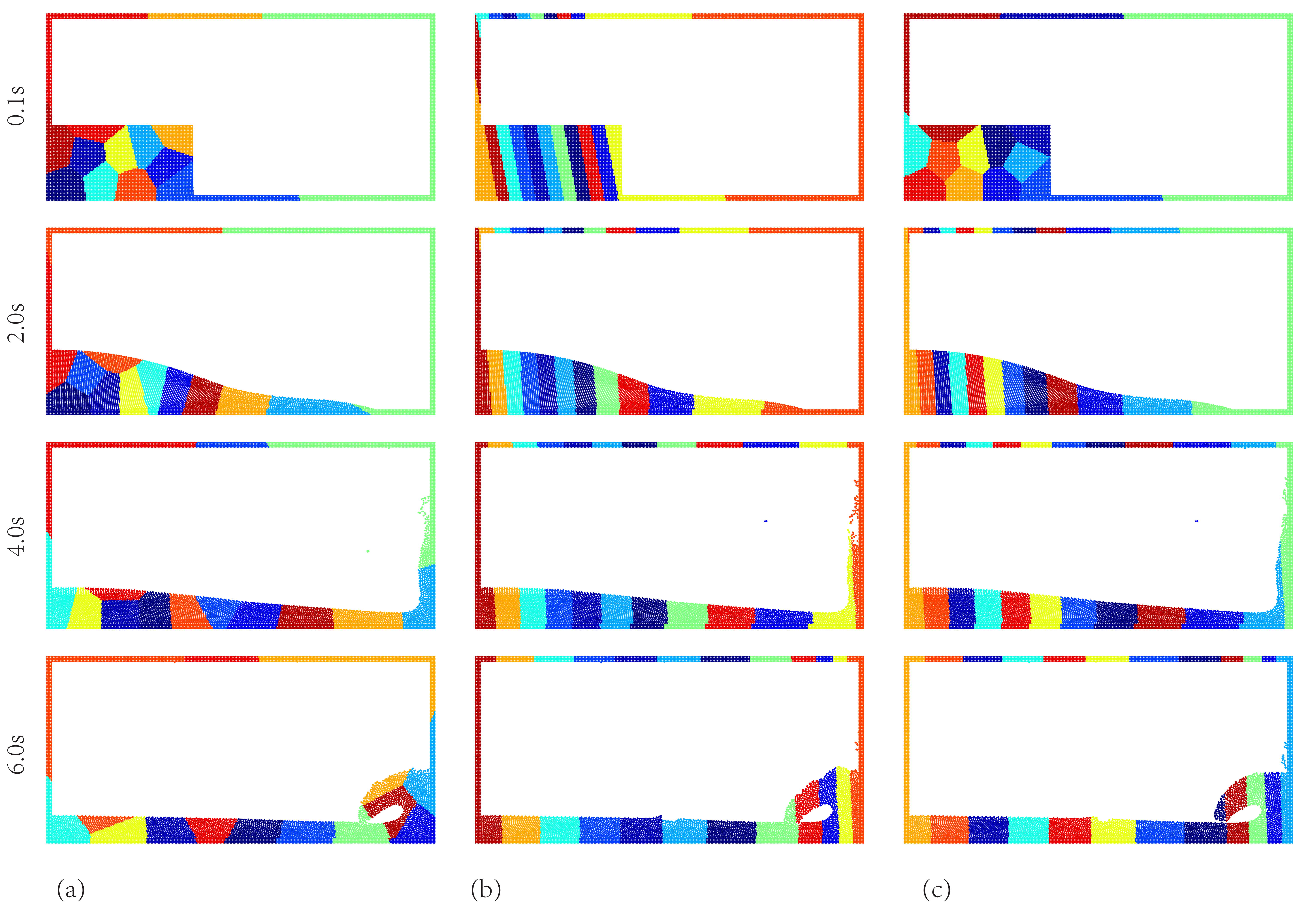}
\caption{2D dam-break: snapshots of simulation result at four instants using (a) original CVP method with background velocity, (b) Inertial CVP with adaptive filter switched off and (c) Inertial CVP with adaptive filter.}
\label{Fig:S5_dambreak_color}
\end{figure}

\begin{figure}[h!]
\centering
\includegraphics[width=0.8\textwidth]{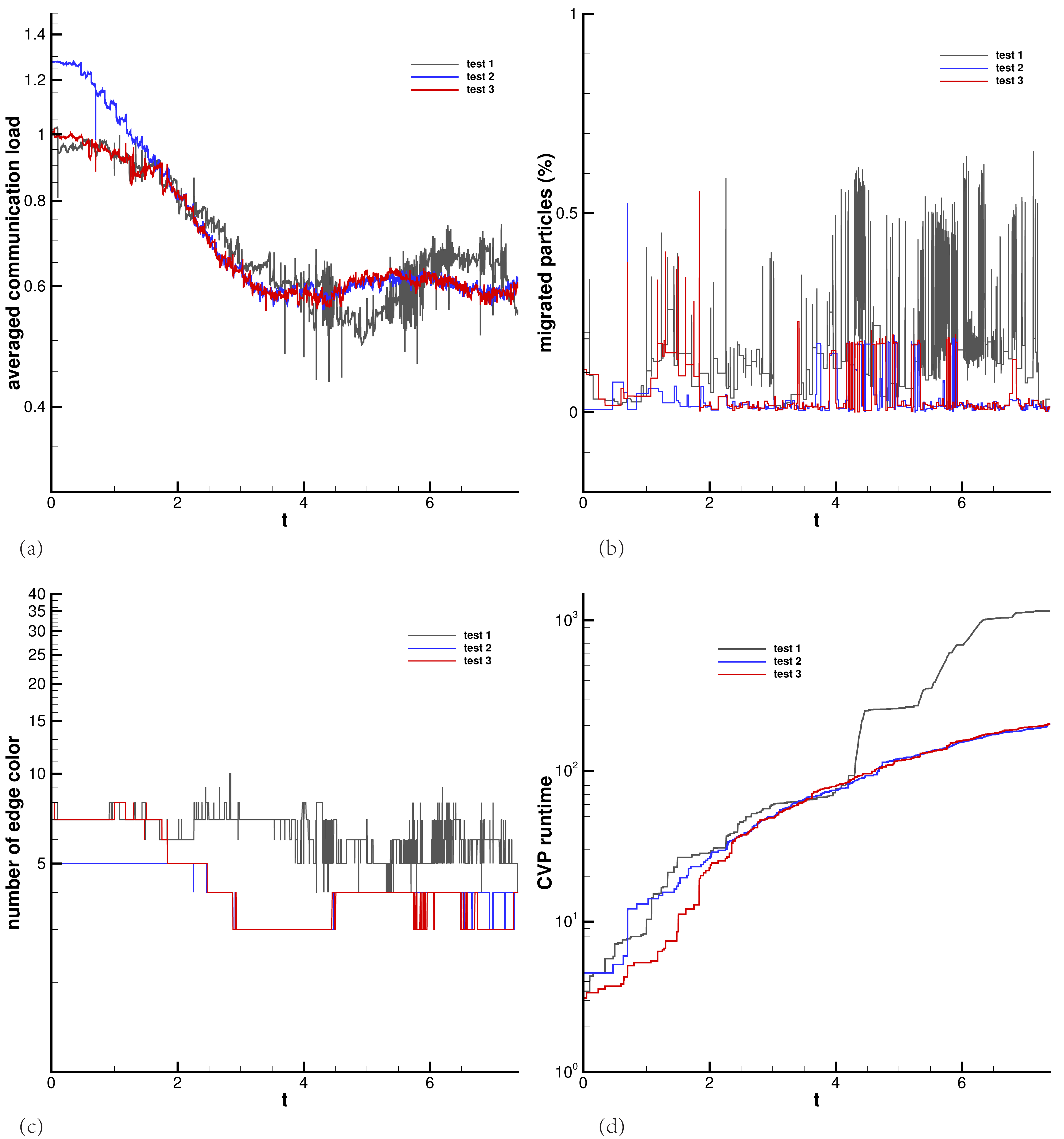}
\caption{2D dam-break: statistic comparison of original CVP method with background velocity, Inertial CVP with adaptive filter switched off and on. (a) time history of averaged communication load, i.e. proportion of ghost buffer particles versus local registered particles. (b) time history of averaged migrated particles, i.e. proportion of migrated particles versus local registered particles. (c) time history of edge color number. (d) runtime regarding to domain decomposition subroutine, i.e. CVP method and data migration.}
\label{Fig:S5_dambreak}
\end{figure}

Finally, we consider the 2D dambreak case following the setup from \citep{adami2012generalized}. Initially, a liquid column is placed at the corner of the sink, and collapses due to the existence of gravity. The evolution of flow consists of violent wave-breaking and splashing event, witch is crucial in free-surface flow modeling. Meanwhile, the dynamic characteristic of this case deteriorates the performance of repartitioner as well, as the computational load varies dramatically in space and rebalancing becomes expansive. Three means are employed here for a comprehensive assessment of the performance: (1) original CVP method with background velocity; (2) proposed LICVP method with adaptive filter switched off, i.e. only one constraint is applied throughout the simulation, and here only partitioning along the largest principle inertial axis is allowed; (3) proposed LICVP method with adaptive filter, and we set $ \lambda_{max}=0.81 $ and $ \lambda_{min}=0.19 $. The background velocity is calculated by Eq. \ref{eq:VP_backgroud_velocity_1}. The number of particles simulated is 8208, and 12 MPI tasks are launched.

Fig \ref{Fig:S5_dambreak_color} (a),(b) and (c) manifests the simulation result at 4 instants regarding to test 1, 2 and 3 respectively. Test 1 and 3 exhibits similar partitioning result at $ 0.1s $, when the water column is not sufficiently influenced by gravity. Whereas test 2 features sub-domains of slim-shaped rectangular, since only one dof (degree of freedom) is unconstrained. Following the propagation of wave front, the load distribution of the system varies accordingly and the load along horizontal axis becomes dominant. At $ t=2s $, the partitioning strategy in test 3 is altered, and constraint is applied along $ \textbf{N} $, i.e. identical to test 2. The partitioning strategy remains the same afterwards. It is observed that, the incremental property is best preserved in test 2, where the topology of sub-domains is identical during entire simulation. The incremental property for test 1 is the worst, as the topological layout of sub-domains varies constantly. Test 3 generally achieves an intermediate performance, since the partitioning strategy is dynamically shifted at $ t=2s $.

The communication load is compared in Fig. \ref{Fig:S5_dambreak} (a). $ S_c $ for test 2 is the largest before $ t=2s $, while test 1 and 3 has close communication volume during this period. The slim-shaped sub-domains cannot guarantee the optimization of communication reduction comparing to the compact sub-domains presented in test 1 and 3. After $2s$, test 2 and 3 exhibit generally the same level of communication load, whereas test 1 achieves lower $ S_c $ during $ 4s $ to $ 6s $ and surpasses test 2 and 3 afterwards. Although test 1 achieves roughly equivalent performance in optimizing communication load after $2s$, the number of communication sub-steps, i.e. edge color, is higher and the system is rebalanced more frequently (see Fig. \ref{Fig:S5_dambreak} (c)). Regarding to incremental property, the same conclusion can be drawn as mentioned in last paragraph by comparing the averaged data migration (see Fig. \ref{Fig:S5_dambreak} (b)). The benefit from switching partitioning strategy in test 3 is remarkable, which avoids the violent stage encountered with original CVP method. The runtime comparison is illustrated in Fig. \ref{Fig:S5_dambreak} (d)). Test 1 is the most time consuming, and with the proposed Inertial CVP method, a speedup of about 6x is achieved.

\section{Conclusions}
\label{S:6}

In this paper, a Lagrangian Inertial CVP method is developed by compounding the concepts of a background velocity as well as an Inertial CVP method. The proposed LICVP method is employed as the rebalancer of a multi-resolution parallel framework for SPH method. The main accomplishment can be summarized as follows:

\paragraph{1}
By defining a background velocity, Voronoi particles are able to track the motion of local sub-domains and characterize the topological variation of the system more precisely. Rebalancing upon the updated Voronoi-particle positions improves the incremental property remarkably. Moreover, since the equilibrium is calculated globally, the inter-processor communication is reduced implicitly.

\paragraph{2}
The performance of simulations with extremely anisotropic computation-load distribution is improved utilizing the proposed Inertial CVP method. Due to the splitting operator, the Voronoi-particle motion is insensitive of the load variation along directions of minimum interest, which enhances the incremental property, and improves the convergence as well. Additionally, the adaptive filter allows a dynamic selection of partitioning strategy according to the evolution of load distribution. The selection procedure guarantees a relative balance between data-redistribution and inter-processor communication cost in extreme situations.





\section*{Acknowledgements}
The first author is partially supported by China Scholarship Council (NO. 201506290038). The second author is partially supported by China Scholarship Council (NO. 201206290022).The computational resources are provided by Leibniz- Rechenzentrum der Bayerischen Akademie der Wissenschaften, München (LRZ).
%



  \bibliographystyle{elsarticle-num}
  \scriptsize
  \setlength{\bibsep}{0.5ex}

\bibliography{MRSPH_02}

\end{document}